\newif\ifmnras
\mnrastrue
\ifmnras
    	\documentclass[apj,numberedappendix]{emulateapj}
\else
	\documentclass[apj,numberedappendix]{emulateapj}
\fi

\usepackage{graphics,epsf}
\usepackage{amsmath}                
\usepackage{amsfonts}               
\usepackage{amssymb}                
\usepackage{epsfig}                 
\usepackage{float}
\usepackage{color}

\def \eV{~\rm{eV}}

\def \s{~\rm{s}}
\def \km{~\rm{km}}

\def \g{~\rm{g}}

\def \erg{~\rm{erg}}

\def \yr{~\rm{yr}}

\ifmnras
	\def \aap{A\&A}
	
	\def \apj{ApJ}
	\def \apjl{ApJ}
	\def \apjs{ApJS}

	\def \mnras{MNRAS}
	\def \na{New Astron.}
    \def \nar{New Astron. Rev}
    
	\def \pasa{Publ. Astron. Soc. Australia}

\fi

\usepackage{xcolor}
\definecolor{redak}{rgb}{0.9,0.15,0.05}

\begin{document}

\ifmnras

\title{The limited role of recombination energy in common envelope removal}
\author{Aldana Grichener\altaffilmark{1}, Efrat Sabach\altaffilmark{1} \& Noam Soker\altaffilmark{1,2}}

\altaffiltext{1}{Department of Physics, Technion -- Israel Institute of Technology, Haifa 32000, Israel; aldanag@campus.technion.ac.il; efrats@physics.technion.ac.il; soker@physics.technion.ac.il}
\altaffiltext{2}{Guangdong Technion Israel Institute of Technology, Shantou, Guangdong Province, China}
 
\begin{abstract}
We calculate the outward energy transport time by convection and photon diffusion in an inflated common envelope and find this time to be shorter than the envelope expansion time. We conclude therefore that most of the hydrogen recombination energy ends in radiation rather than in kinetic energy of the outflowing envelope. We use the stellar evolution code \texttt{MESA} and inject energy inside the envelope of an asymptotic giant branch star to mimic energy deposition by a spiraling-in stellar companion. During 1.7 years the envelope expands by a factor of more than 2. Along the entire evolution the convection can carry the energy very efficiently outwards, to the radius where radiative transfer becomes more efficient. The total energy transport time stays within several months, shorter than the dynamical time of the envelope. Had we included rapid mass loss, as is expected in the common envelope evolution, the energy transport time would have been even shorter. It seems that calculations that assume that most of the recombination energy ends in the outflowing gas might be inaccurate. 
\end{abstract}



\section{INTRODUCTION}
\label{sec:intro}

Thirty years of three-dimensional (3D) hydrodynamical simulations of the common envelope evolution (CEE) have consistently showed that the envelope inflates and that mass ejection is concentrated toward the equatorial plane (e.g., \citealt{LivioSoker1988, RasioLivio1996, SandquistTaam1998, Sandquistetal2000, Lombardi2006, RickerTaam2008, TaamRicker2010, DeMarcoetal2011, Passyetal2011, Passyetal2012, RickerTaam2012, Nandezetal2014, Ohlmannetal2016, Ohlmannetal2016b, Staffetal2016MN8, NandezIvanova2016, Kuruwitaetal2016, IvanovaNandez2016, Iaconietal2017, DeMarcoIzzard2017, Galavizetal2017, Iaconietal2017a, MacLeodetal2018}).
These expected outcomes came along with the understanding that it is not so simple to eject the common envelope in a short time, and that in addition to the orbital gravitational energy of the binary system another energy source is required to unbind the entire envelope. This unsuccessful envelope removal is not due to numerical limitations, e.g., of not having convection and radiative transfer or of a limiting resolution, but rather it seems to be a fundamental property of the CEE, likely due to the small envelope mass at short orbital separations \citep{Soker2013}.  
In what follows we refer to the giant star as the primary, and to the more compact companion as the secondary star. 

Although the gravitational energy that is released during the in-spiral of the binary system is larger than the binding energy of the common envelope (CE; e.g., \citealt{DeMarcoetal2011, NordhausSpiegel2013}), other energy sources should play a non-negligible role. The extra energy source can act on a short time scale at the beginning of the CE, or can be effective over a long time, up to the ejection of the entire envelope in the last phase of the CEE. 

In a recent paper \cite{Soker2017final} summarizes these extra energy sources.  
The processes that might contribute energy over a long time in the last CE phase are excitation of p-waves in the giant envelope \citep{Soker1992, Soker1993}, enhanced dust formation above the inflated and rotating envelope \citep{Soker2004, GlanzPerets2018},
interaction of the core-secondary system with a circumbinary disk 
at the final phase of the CEE (e.g., \citealt{Kashisoker2011MN, Kuruwitaetal2016}),  envelope inflation followed by vigorous pulsation \citep{Claytonetal2017}, and jets launched by the secondary star as it accretes mass from the circumbinary envelope when it is about to exit the CE from inside \citep{Soker2017final}. 

The processes that in principle are thought to release extra energy on a relatively short time scale at the beginning of the CEE, i.e., mainly during the plunge-in phase, are jets and recombination energy. 
  
The jets that are launched by the secondary star, being a main sequence star or a more compact object, interact hydrodynamically with the envelope and facilitate the removal of the envelope (see \citealt{Soker2016Rev} for a review and, e.g., \citealt{MorenoMendezetal2017} for recent hydrodynamical simulations). In some cases the jets can efficiently remove the outer envelope zone when the secondary star approaches and just enters the envelope to prevent the CEE. 
The secondary star continues to graze the giant envelope as it removes all the envelope mass outside its orbit. The system avoids, at least for a while, the CEE. This type of evolution is termed the Grazing envelope evolution (GEE; e.g., 
\citealt{Soker2015, Shiberetal2016, ShiberSoker2018, AbuBackeretal2018}). The likely formation of an accretion disk around the secondary star in the outer zones of the envelope \citep{MurguiaBerthieretal2017} suggests that jets can be launched as the secondary grazes or just enters the envelope. 
In the inner zones of the envelope the formation of an accretion disk is more difficult (e.g., \citealt{MacLeodRamirezRuiz2015disk}; but see disk formation around a neutron star companion, \citealt{MacLeodRamirezRuiz2015NS}), but possible if energy removal by the jets themselves is considered \citep{Shiberetal2016}. 

In this paper we do not consider the pre-CEE mass loss that can take place due to tidal spin-up (e.g., \citealt{BearSoker2010}), mass transfer (e.g., \citealt{MacLeodetal2018}), or due to the grazing envelope evolution, and in some cases might be quite substantial. 

The other extra energy source that has been proposed to facilitate common envelope removal is the recombination energy of hydrogen and helium. 
(e.g., \citealt{Nandezetal2015, IvanovaNandez2016, Kruckowetal2016, NandezIvanova2016} for recent papers).
However, in the outer envelope the hydrogen is mostly neutral and the optical depth is low. Consequently photon outward diffusion time is short, and the efficiency of utilizing the recombination energy to envelope removal turns out to be low \citep{Harpaz1998, SokerHarpaz2003}. In a recent study \cite{Sabachetal2017} find in their 1D stellar simulations that not only the photon diffusion time is short, but also convective time is short. Therefore they conclude that recombination energy cannot be the main extra energy source of common envelope removal.
In addition, the recombination energy is a tiny fraction of the energy that is needed to eject the envelope from sub-giant branch primary stars, and hence cannot play a significant role in those cases of envelope ejection. For example, in the stellar merger model presented by \cite{MacLeodetal2017M31} for the transient M31LRN 2015 in the Andromeda galaxy, the recombination energy of the ejected gas is   $< 2 \%$ of the kinetic energy of the ejecta. 

\cite{Sabachetal2017} did not consider in their calculations the expansion of the envelope during the early CEE. In the present study we follow the early inflation of the common envelope, to about 2.1 times its initial radius, and calculate the energy removal time by radiation and convection. 
We differ from earlier calculations of the energy transport by convection 
(e.g., \citealt{MeyerHofmeister1979, Podsiadlowski2001, Ivanovaetal2015}) 
as these study the convection during the later and longer CE phases, while we study the short plunge-in phase. In the long duration phase of the CEE convection has enough time to transfer the recombination energy out. Hence we concentrate on the early phase where there are more open questions regarding energy transport.  

We describe the basic energy transport time scales in section \ref{sec:time}.
We then describe the numerical procedure in section \ref{sec:scheme}, the stellar evolution in section 
\ref{sec:evolution} and the results in section \ref{sec:results}. Our summary is in section \ref{sec:summary}. 

\section{Energy lose time scales}
\label{sec:time}

We take the photon diffusion time out of a recombination zone at a depth of $\Delta R$ below the photosphere to be as in \cite{Sabachetal2017}  
\begin{equation}
t_{\rm diffusion} \approx \frac{ 3 \tau \Delta R}{c} =
5.5 
 \left( \frac{\tau}{2.5\times 10^5} \right)
  \left( \frac{\Delta R}{100 R_\odot} \right)
\label{eq:tdiff} \yr,
\end{equation}
where $\tau$ is the optical depth from the recombination zone outwards, and $c$ is the speed of light.
We scaled the values of $\tau$ and $\Delta R$ with the typical quantities of the zone where the ionization fraction of hydrogen is $50 \%$ in our AGB model after substantial expansion (see section \ref{sec:scheme} for the stellar model). 

We consider the relatively short time during which the secondary star dives from the surface of the giant and deep into the envelope. Not only is the envelope convective to start with before the CEE, but \cite{Ohlmannetal2016} find in their 3D hydrodynamical simulations that instabilities during the CEE indicate the onset of turbulent convection.
They mention that convection can play a significant role in the energy transport on thermal time scales. \cite{Sabachetal2017} deduce that convection can transport energy on a dynamical time scale. We here find that the convection might be even more efficient than what they assume.  

For the maximum power of energy transport by subsonic convection \cite{Sabachetal2017} adopt the expression given by \cite{QuataertShiode2012} for regular convection and write 
$ L_{\rm max,conv}(r) = 4 \pi \rho(r) r^2 c_s^3(r),$ 
where $\rho (r)$ and $c_s(r)$ are the density and the sound speed at radius $r$,
respectively. 
This expression takes the heat per unit mass content of the gas to be $c_s^2$. 
In our stellar model (section \ref{sec:scheme}) the sound speed in the $50 \%$ ionization level of hydrogen is $c_s ({\rm H}^+) \simeq 11 \km \s^{-1}$. 
At $90 \%$  and $10 \%$ hydrogen ionization fractions the sound speeds are about $14 \km \s^{-1}$ and about $8 \km \s^{-1}$, respectively. 

It turns out that the specific heat content of the gas can be larger on average than the sound speed. For the recombination energy not to escape the diffusion time should be long. Hence, the mass parcel retains its recombination energy for a dynamical time, during which the parcel of gas can move outwards. For a solar composition the specific heat of the gas corresponds to a specific energy of 
$e_{\rm rec}({\rm H}^+)= 13.6 X_{\rm H} \eV /m_H=9 \times 10^{12} \erg \g ^{-1} = (30 \km \s^{-1})^2$, 
where $X_{\rm H}$ is the mass fraction of hydrogen and $m_H$ is the hydrogen atomic mass. 
Namely, the heat content of the parcel of gas as it recombines is larger by a factor of few than the simple estimate of  $c^2_s ({\rm H}^+)$.  
Even if we take half the total recombination energy at $50 \%$ ionization, we still find larger energy transport. We therefore take the maximum possible convection energy transport to be 
\begin{equation}
L_{\rm max,conv}({\rm H}^+) \simeq \beta 4 \pi \rho(r) r^2 c_s^3(r),               \label{eq:lmaxconv1}
\end{equation}
where $\beta \simeq 5$. 

In other words, the convective cell can carry most of the recombination energy outwards on a time scale of 
\begin{eqnarray}
\begin{aligned}
t_{\rm conv, min} \simeq & \int_{r({\rm H}^+)}^R  \frac{dr}{c_s(r)}
\\ & \simeq 0.22 
\left[ \frac{\Delta R({\rm H}^{+})}{100 R_\odot} \right]
\left( \frac{\bar c_s}{10 \km \s^{-1}} \right)^{-1} \yr,
\end{aligned}
\label{eq:Tout}
\end{eqnarray}
where $\bar c_s$ is the average sound speed that is the maximum convection speed, $R$ is the photosphere radius, and $\Delta R({\rm H}^{+}) =R-{r({\rm H}^+)}$ is the depth of the $50 \%$ ionization zone of hydrogen. 
   
Later on we will calculate the actual convective time by considering the actual convective velocity $v_{\rm conv} (r)$ which is in most cases lower than the sound speed
\begin{equation}
t_{\rm convection} (r) = \int_{r}^R  \frac{dr}{v_{\rm conv}(r)}.
\label{eq:Tconv}
\end{equation}

\section {The numerical scheme}
\label{sec:scheme}
We run the stellar evolution code \texttt{MESA} (Modules for Experiments in Stellar Astrophysics), version 9575 (\citealt{Paxtonetal2011, Paxtonetal2013, Paxtonetal2015,Paxtonetal2018}) to study the typical time scales of photon diffusion and convection in an asymptotic giant branch (AGB) star. \cite{Sabachetal2017} calculated these quantities for an AGB star with an initial mass (on the zero age main sequence; ZAMS) of $M_{\rm 1,ZAMS}=2 M_{\odot}$ with a metallicity $Z=0.02$. However, they did not take into account the energy deposited to the envelope by the in-spiraling secondary star in a CEE. We here examine the effect of this energy.
 
As in \cite{Sabachetal2017} we take a star with a ZAMS mass of $M_{\rm 1,ZAMS}=2 M_{\odot}$ and metallicity $Z=0.02$. We start the energy injection when the primary star is an AGB star with a mass of $M_{\rm 1,AGB}=1.75 M_{\odot}$ and a radius of $R(t=0) = 250R_\odot$.

To enable the calculation with a 1D stellar code, we make the following simplifying assumptions.
\begin{enumerate}
\item We mimic the spiraling-in of a secondary star by simply depositing energy into the AGB stellar envelope. We take the energy and time scale of energy deposition following the study of \cite{Passyetal2012} for a companion of mass $M_{\rm 2}=0.3 M_{\odot}$. 
\item We do not include the rapid mass loss expected as the secondary spirals in. 
\item We neglect the possibility that the secondary star accretes mass and launches jets. 
\item We neglect the rotation of the envelope.
\end{enumerate}
 
Let us elaborate on the energy injection scheme. 
To prevent numerical problems and to maintain the number of parameters as small as possible, we deposit energy in a constant radial range inside the envelope,
and avoid the outer regions. In any case, the deposition of gravitational energy increases as orbital separation decreases, and so even in an accurate calculation most of the energy is deposited in the inner regions of the envelope. 
The energy is injected at a constant power during the brief initial stage of CEE. According to \cite{Passyetal2012}, the separation between the primary core and the companion at the end of the rapid in-fall (plunge-in) phase is $\approx 0.2 R(t=0)$. We take the plunge-in to end at a radius of $0.2R=50R_\odot$ and deposit the energy in inner numerical shells that satisfy $50 R_{\odot}<r< 120 R_{\odot}$, where the upper bound is somewhat arbitrary.
At each time step we distribute the energy in the injection zone with a constant energy per unit mass, $q \Delta t/m_{\rm in}$, where $q$ is the power, $\Delta t$ is the time step, and $m_{\rm in}(t)$ is the mass inside the injection zone.

The exact place and manner by which energy is injected into the envelope change the exact outcome of the CEE. For example, a more massive companion that ends at a larger orbital separation will deposit the same amount of energy to the envelope yet the result will be different. In the present study we are interested only in examining how the energy transport time from the ionization zone of hydrogen varies as the envelope is inflated. Because the ionization zone is in the outer regions of the envelope, our calculations depend weakly on the response of the inner regions of the envelope where we deposit the energy. 
For our purposes, therefore, the exact numerical scheme of energy deposition that takes place in the envelope below the ionization zone of hydrogen is not important, as long as we achieve envelope inflation. 

We estimate the plunge-in time of the companion according to the work of \cite{Passyetal2012}, where the plunge-in time of a secondary star of $M_{\rm 2}=0.3 M_{\odot}$ orbiting around a primary star with an RGB mass of $M_{\rm 1,RGB}=0.88 M_{\odot}$ is 280 days.
If we take the ratio of the plunge-in time to dynamical time on the giant surface in our model to be as in their case, the plunge-in time would be $2.9 \yr$.
Considering that our envelope is three times more massive than their model, we take the plunge-in time to be somewhat shorter, $t_{\rm plunge-in}=1.7 \yr$. 
As we discuss later on, our conclusions hold even for a half as short plunge-in time of about 10 months instead of 20 months (1.7 year). 

Since we neglect the accretion process on to the secondary star and the rotation of the envelope, the energy we deposit into the envelope is the gravitational energy released by a secondary star of mass $M_{\rm 2}=0.3 M_{\odot}$ that spirals-in from the AGB surface at an orbital separation of $a_0=R(t=0)=250 R_\odot$ to a core-secondary orbital separation of $a_f=50 R_\odot$ and with a core mass of $M_{\rm core}=0.56 M_\odot$. For all these, we take the energy that is channeled to heat within the 1.7 years to be 
$\Delta E_{\rm heat}=(G M_2 M_{\rm core} / 2 a_f)-(G M_2 M_{\rm 1,AGB} / 2 a_0) 
= 2.4 \times 10^{45} \erg$.
Over all, we inject energy with a power of 
\begin{equation}
q = \frac{\Delta E_{\rm heat}}{t_{\rm plunge-in}} = \frac{2.4 \times 10^{45} \erg}{1.7 \yr}
= 4.5 \times 10^{37} \; {\rm \erg\; s^{-1}}.
\label{eq:Epower}
\end{equation}
 
After 1.7 years the stellar radius has grown by a factor of about 2 (see below). We do not continue beyond this time as then other effects, such as mass loss, must be taken into account. Nonetheless, the evolution of the 1.7 years period is sufficient for us to draw conclusions regarding the energy transport time during the plunge-in phase. 
 
\section{Stellar evolution}
\label{sec:evolution}

We start by describing the general evolution of the star during the $1.7 \yr$ of plunge-in when we deposit energy to the envelope according to equation (\ref{eq:Epower}).  

In Fig. \ref{fig:tauLR} we present the evolution of 4 stellar parameters that are relevant to our study. As expected in the CEE, the luminosity of the giant star increases significantly as orbital energy is deposited into the envelope.  The stellar radius and the radius of $50 \%$ hydrogen ionization, $R_{\rm ion,50}$, increase monotonically. In the upper panel we present the optical depth from $R_{\rm ion,50}$ to the photosphere. It turns out that the optical depth decreases slowly. The fluctuations in the value of the optical depth are due to the shell numerical structure of the star and the large contribution of the inner shells to the optical depth.
We can notice that the peaks and troughs in the optical depth occur when there are very small wiggles in the exact location of $R_{\rm ion,50}$ (red line in lower panel).
\begin{figure}
\begin{center}
\vspace*{-2.9cm}
\hspace*{-1.3cm}
\includegraphics[width=0.6\textwidth]{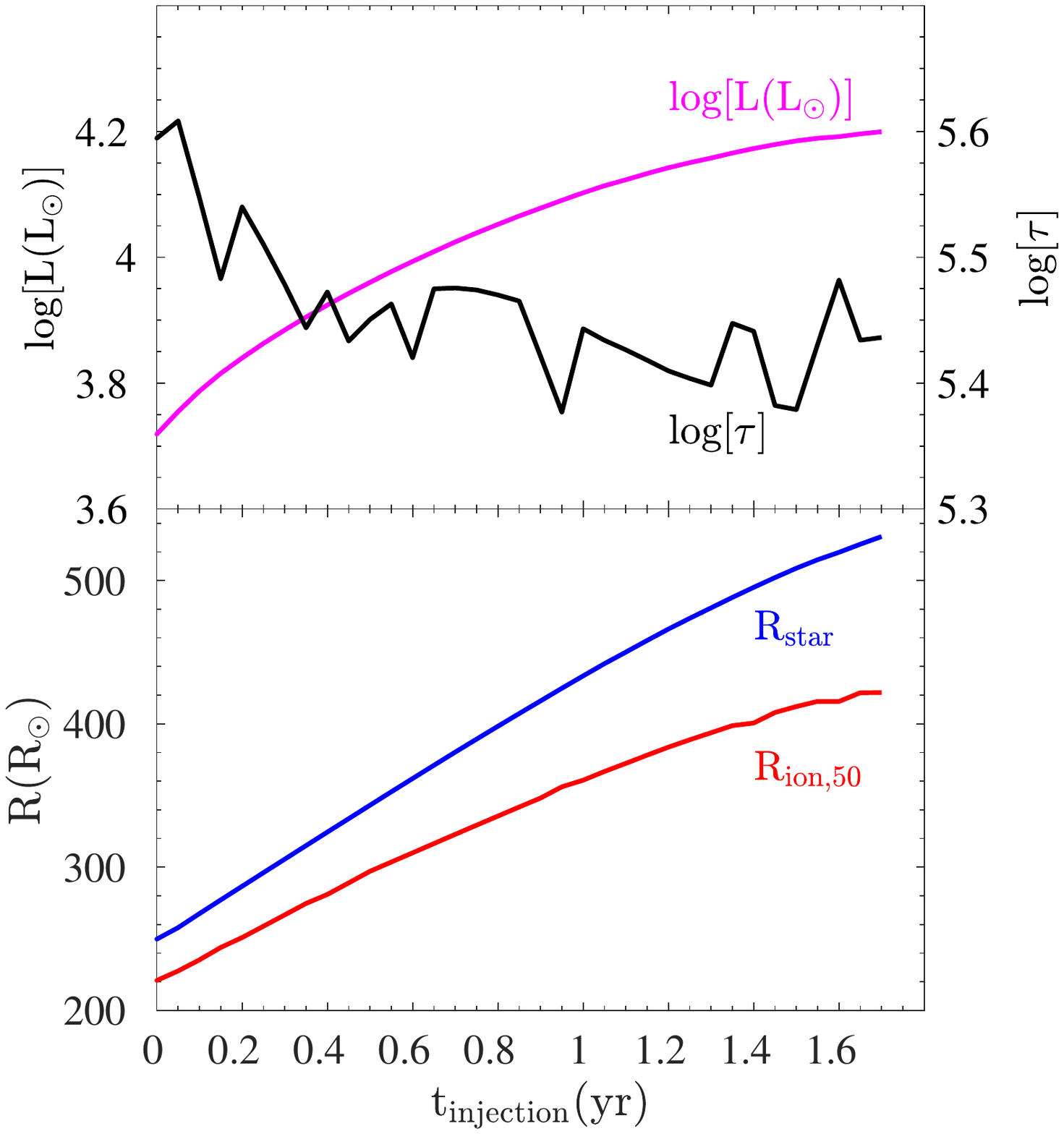}
\vspace*{-3.5cm}
\caption{The evolution of some stellar parameters of the AGB model during energy injection. At this stage the stellar mass is $M_{\rm 1,AGB}=1.75 M_{\odot}$.   
Upper panel: The stellar luminosity (magenta) and the optical depth from the photosphere down to the radius $R_{\rm ion,50}$ where $50 \%$ of the hydrogen is ionized (black). Lower panel: The radius of the photosphere (blue) and of $R_{\rm ion,50}$ (red).  }
\label{fig:tauLR}
\end{center}
 \end{figure}

In Fig. \ref{fig:HR} we present the evolution of the star on the HR diagram from the point where we start the injection of energy until the termination of the calculation, a total of 1.7 years. We mark the time and radius at 5 points.   
\begin{figure}
\begin{center}
\vspace*{-4.5cm}
\hspace*{-1.3cm}
\includegraphics[width=0.6\textwidth]{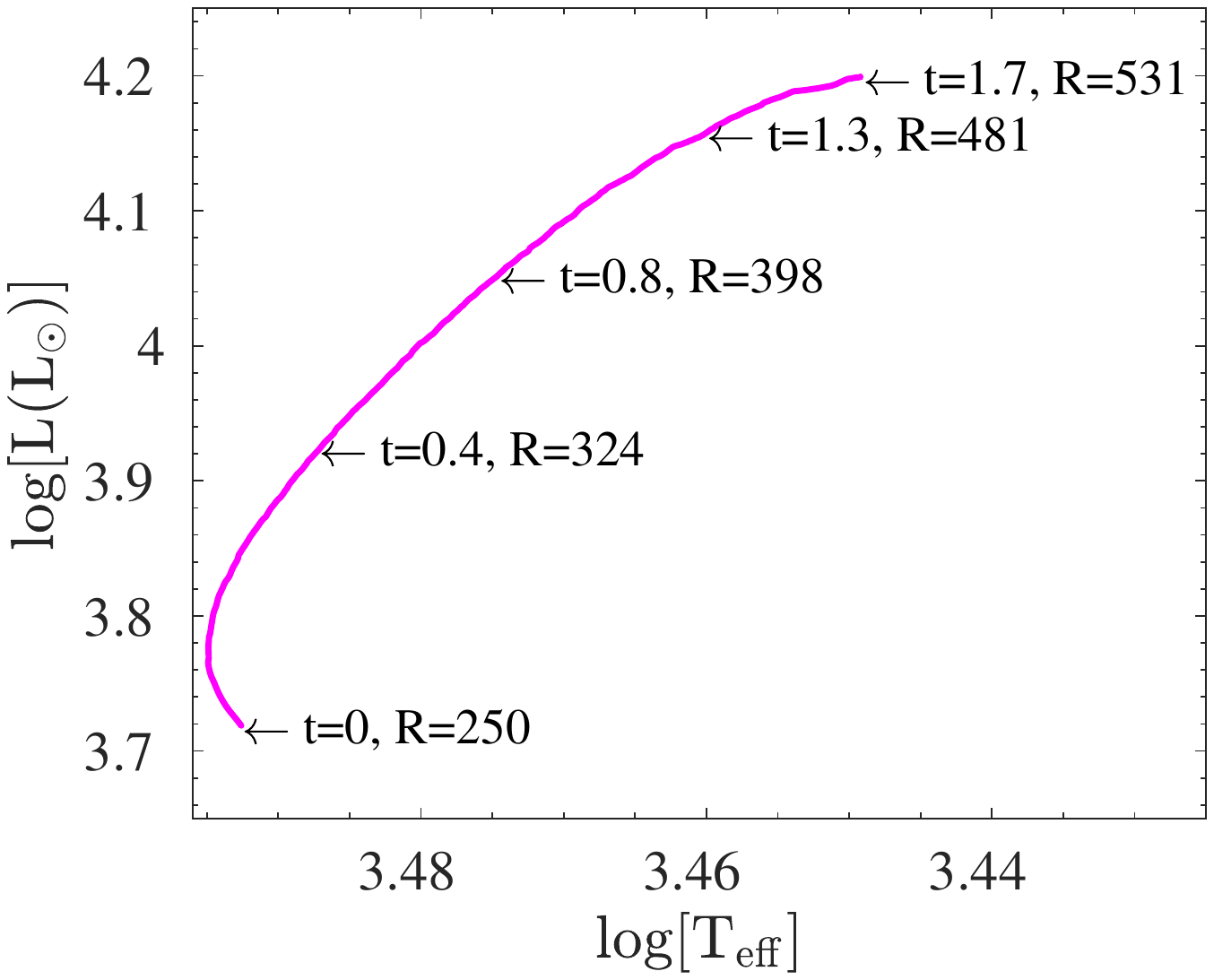}
\vspace*{-4.5cm}
\caption{
The evolution of the AGB star on the HR diagram during our mimicking of the plunge-in phase by energy injection according to equation (\ref{eq:Epower}).  
The numbers near five points on the graph mark the time in years from the beginning of the plunge-in phase, and the stellar radius of the giant in $R_\odot$. 
}   
\label{fig:HR}
\end{center}
\end{figure}

We essentially achieved our goal of mimicking the plunge-in phase by inflating the envelope by a factor of about 2. We now turn to calculate the energy transport time from the ionization zone of hydrogen outwards.  

\section {Diffusion and Convection times}
\label{sec:results}
Our goal is to examine the energy transport time from the recombination zone up to the photosphere as energy deposition inflates the envelope, and compare it with the evolution time, about 1.7 years in the present case. We therefore examine the energy transport time at three evolutionary points, at the beginning of the plunge-in phase, which we take as $t=0$, at $0.8 \yr$, and at the end of the plunge-in phase at $t=1.7\yr$ for our case. 
  
In Figs. \ref{fig:rhoTV0}-\ref{fig:rhoTV17} we present some physical quantities in the outer envelope, where hydrogen is partially ionized, in the three time frames mentioned above. It is important to note that \texttt{MESA} takes into account the recombination energy, and adds it to the envelope. We see that even with the large amount of energy that we add to the envelope as it expands to twice its original radius in a short time, the basic structure stays similar.  
For example, there is a density inversion in the outer part of the envelope and the convection speed stays below the sound speed. This implies that convection can carry even more energy than it carries now. 
\begin{figure}
\begin{center}
\vspace*{-3.13cm}
\hspace*{-1.3cm}
\includegraphics[width=0.6\textwidth]{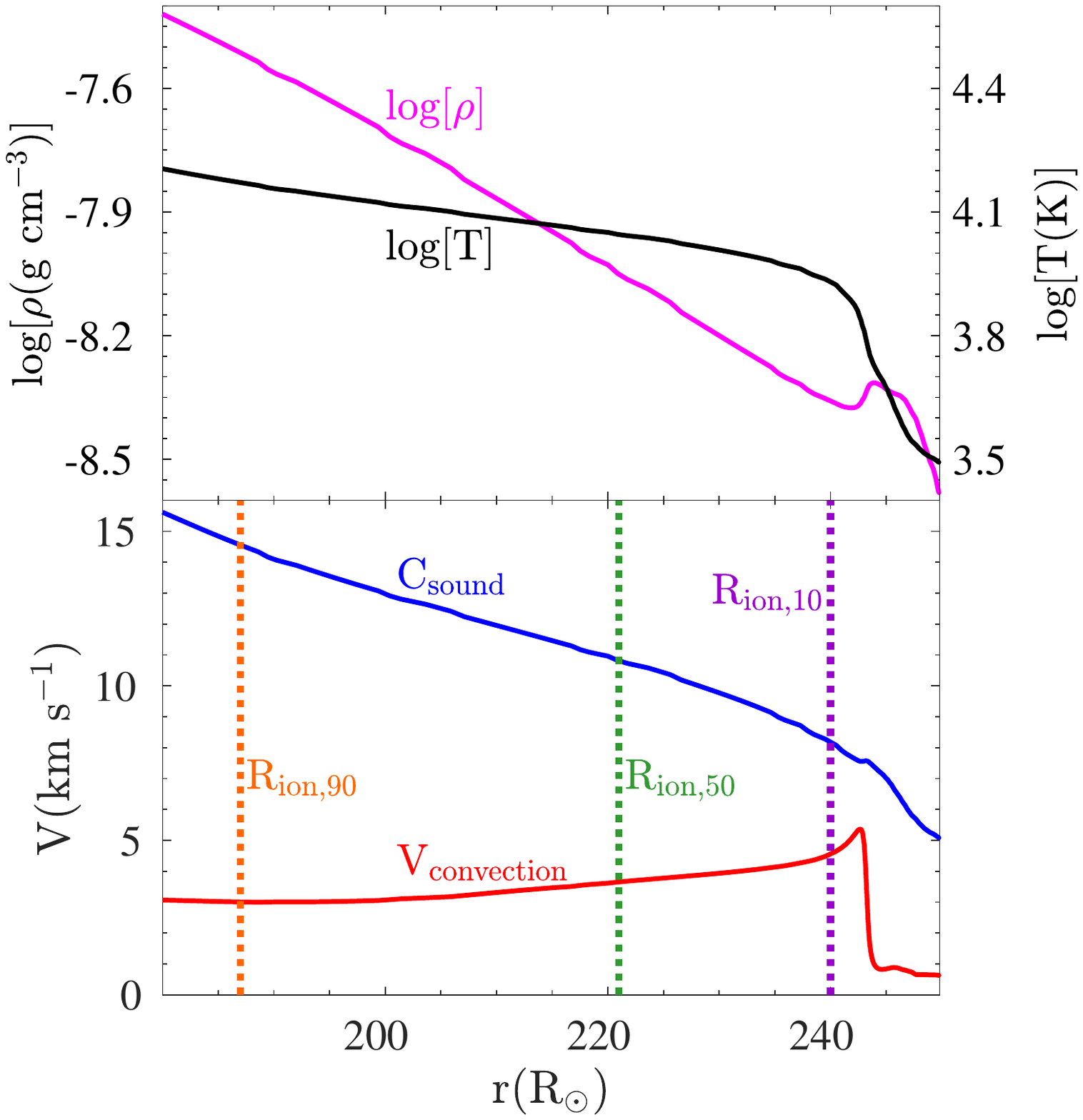}
\vspace*{-3.5cm}
\caption{Profiles of four quantities of our AGB stellar model in the outer layers of the envelope at $t=0$ (the beginning of the plunge-in phase). The right edge of both panels is at the photosphere of the star. Upper panel: The temperature (black) and the density (magenta) in a logarithmic scale. Lower panel: The sound speed (blue) and the convective velocity (red). The orange, green and purple dotted vertical lines mark the zones of $90\%$, $50\%$ and $10\%$ ionization of hydrogen, respectively. 
}
\label{fig:rhoTV0}
\end{center}
\end{figure}
\begin{figure}
\begin{center}
\vspace*{-4.0cm}
\hspace*{-1.3cm}
\includegraphics[width=0.6\textwidth]{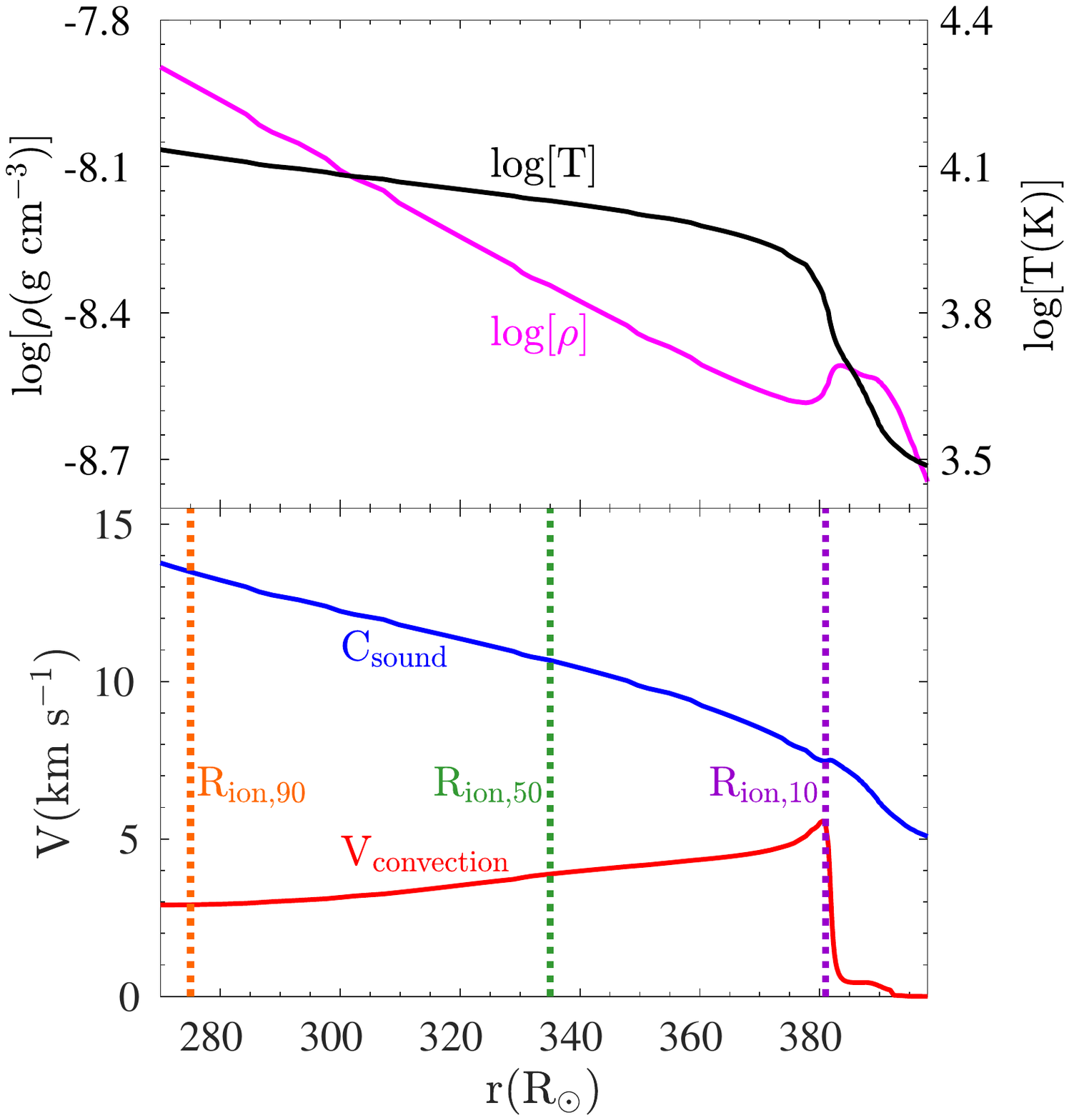}
\vspace*{-3.5cm}
\caption{Like Fig. \ref{fig:rhoTV0} but at $t=0.8
\yr$ from the beginning of the plunge-in phase.         
}
\label{fig:rhoTV08}
\end{center}
\end{figure}
\begin{figure}
\begin{center}
\vspace*{-2.8cm}
\hspace*{-1.3cm}
\includegraphics[width=0.6\textwidth]{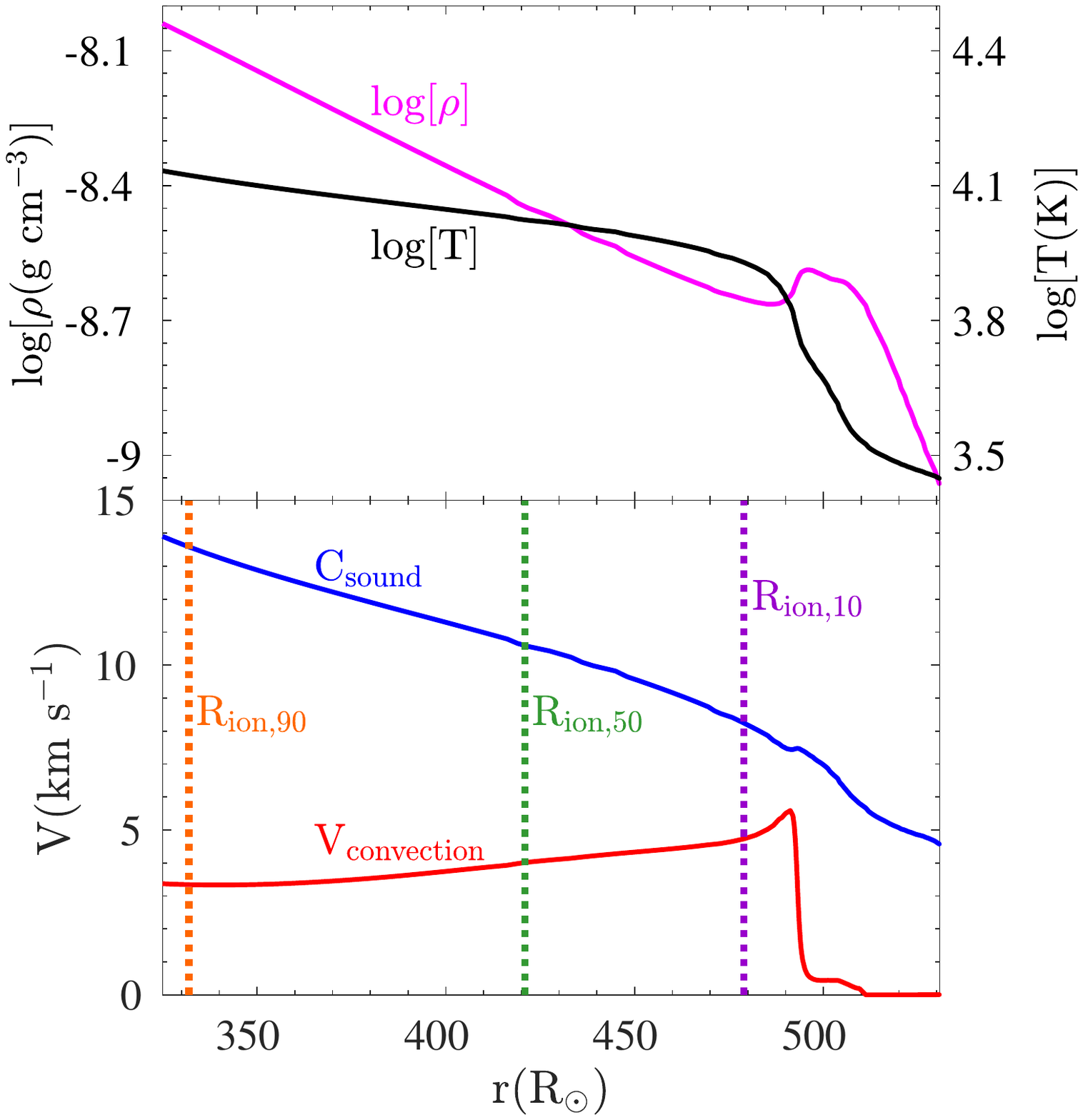}
\vspace*{-3.5cm}
\caption{Like Fig. \ref{fig:rhoTV0} but at the end of the plunge-in phase at  $t=1.7 \yr$.}
\label{fig:rhoTV17}
\end{center}
\end{figure}

We next calculate the energy transport time from radius $r$ to the surface of the star, $R$, by photon diffusion, as given in equation (\ref{eq:tdiff}), and by convection, as given in equation (\ref{eq:Tconv}).  We present these energy transport times at the three evolutionary stages in Figs. \ref{fig:DiffConv0}-\ref{fig:DiffConv1.7}, respectively. 
The convection velocity drops to very low values in the very outer part of the envelope, and so we calculate the convection time to the radius of $1\%$ hydrogen ionization fraction instead of the photosphere. This change is of no significance since in the outer region photon diffusion carries most of the energy.   
\begin{figure}
\begin{center}
\vspace*{-4.5cm}
\hspace*{-1.3cm}
\includegraphics[width=0.6\textwidth]{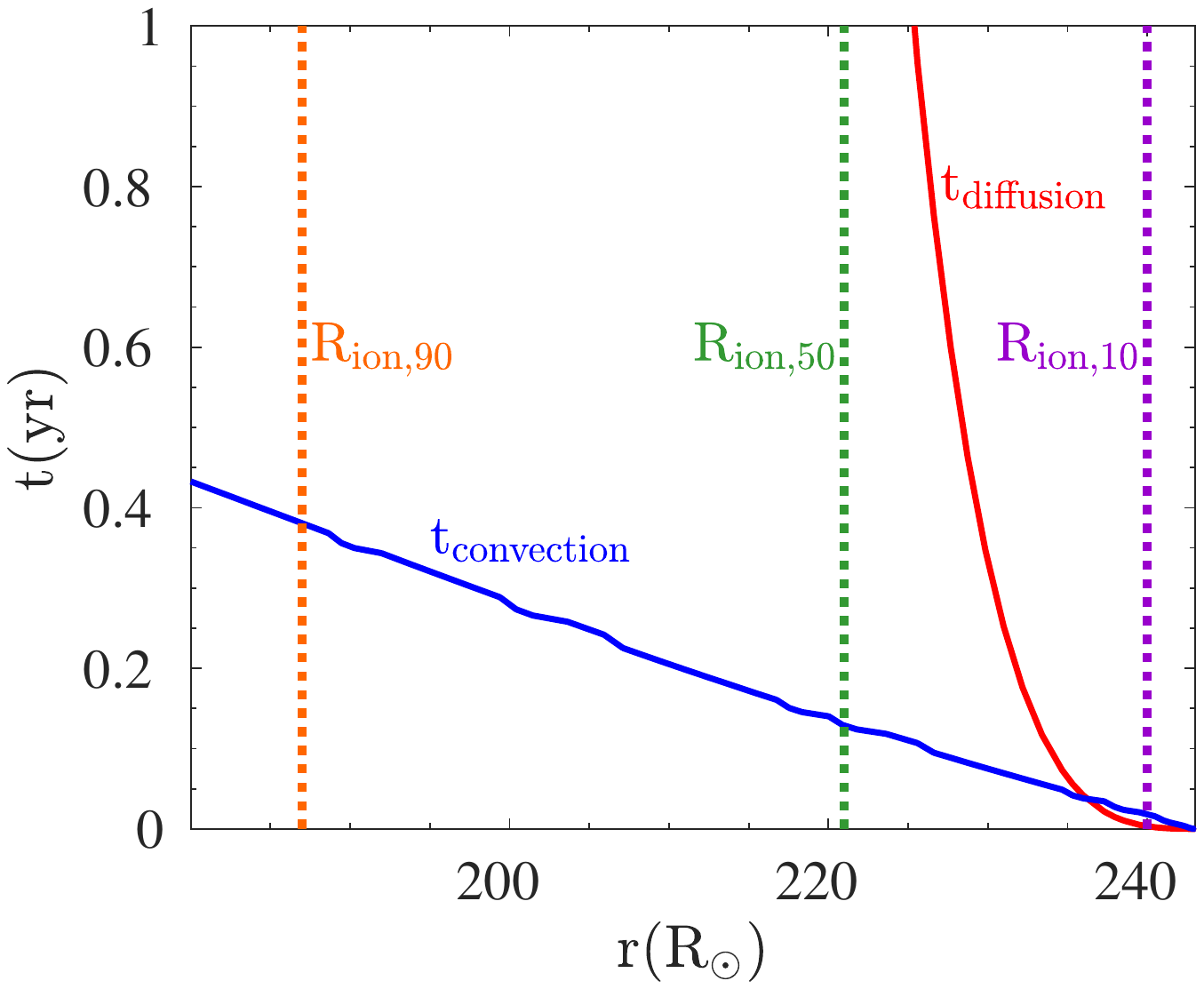}
\vspace*{-4.5cm}
\caption{Energy transport times from radius $r$ to the surface of the star at radius $R$ for $t=0$ when we start the energy injection. Since the convection velocity drops to very low values in the very outer part of the envelope, we calculate the convection time to the radius of $1\%$ hydrogen ionization fraction instead of the photosphere. This radius is also the right boundary of the graph. 
We present the photon diffusion time according to equation (\ref{eq:tdiff}) by the red line, and the convection time according to equation (\ref{eq:Tconv}) with the blue line.
The diffusion time from $R_{\rm ion,50}$ is $t_{\rm diffusion}({\rm H}^+) = 2.2 \yr$.
Note that the photon diffusion time becomes much shorter than the convection time in the outskirts of the envelope. 
}
\label{fig:DiffConv0}
\end{center}
\end{figure}
\begin{figure}
\begin{center}
\vspace*{-4.65cm}
\hspace*{-1.3cm}
\includegraphics[width=0.6\textwidth]{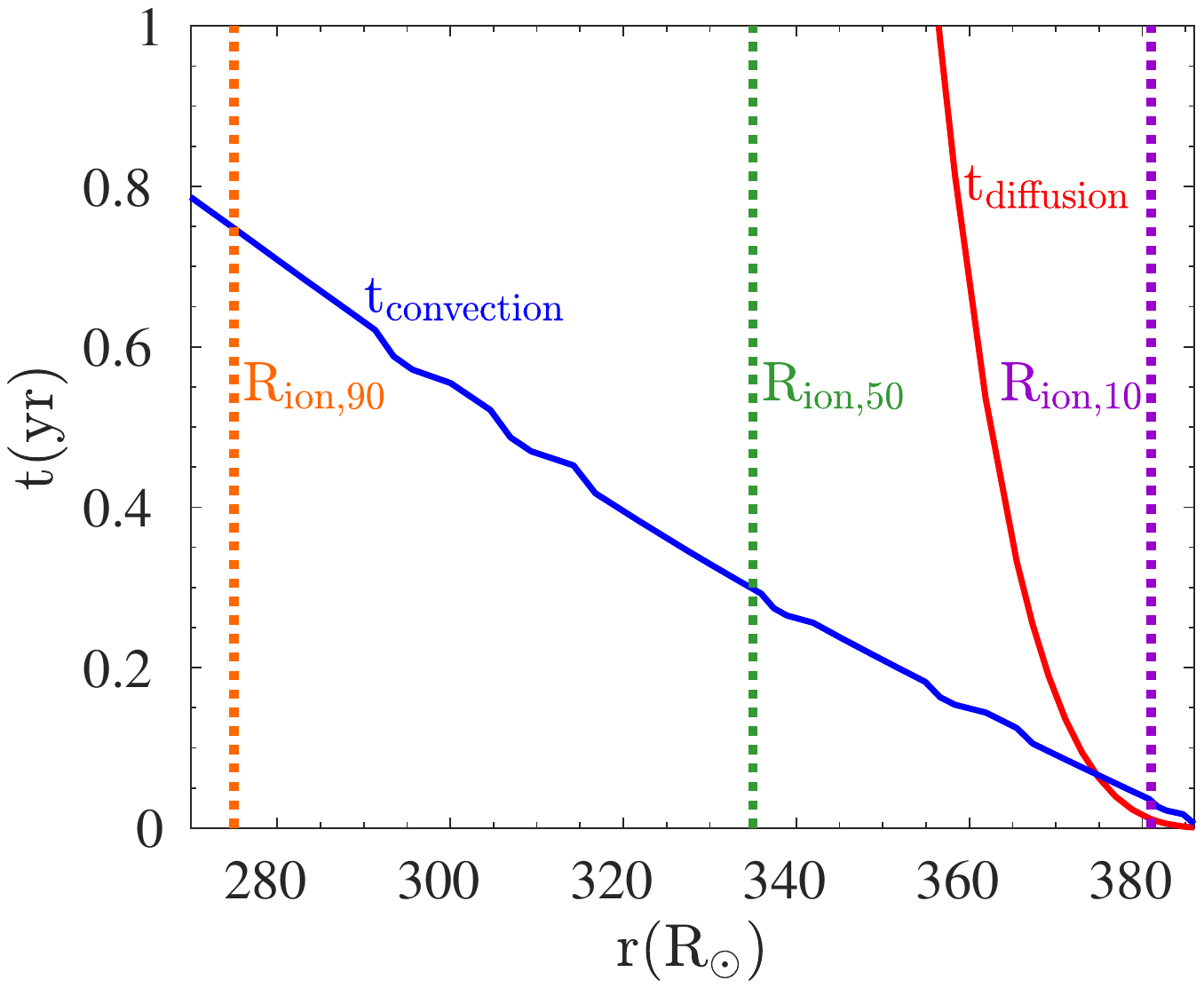}
\vspace*{-4.5cm}
\caption{Like Fig. \ref{fig:DiffConv0} but at $t=0.8  \yr$  from the beginning of the plunge-in phase.
At this time $t_{\rm diffusion}({\rm H}^+) = 3.5 \yr$.
}
\label{fig:DiffConv0.8}
\end{center}
\end{figure}
\begin{figure}
\begin{center}
\vspace*{-4.5cm}
\hspace*{-1.3cm}
\includegraphics[width=0.6\textwidth]{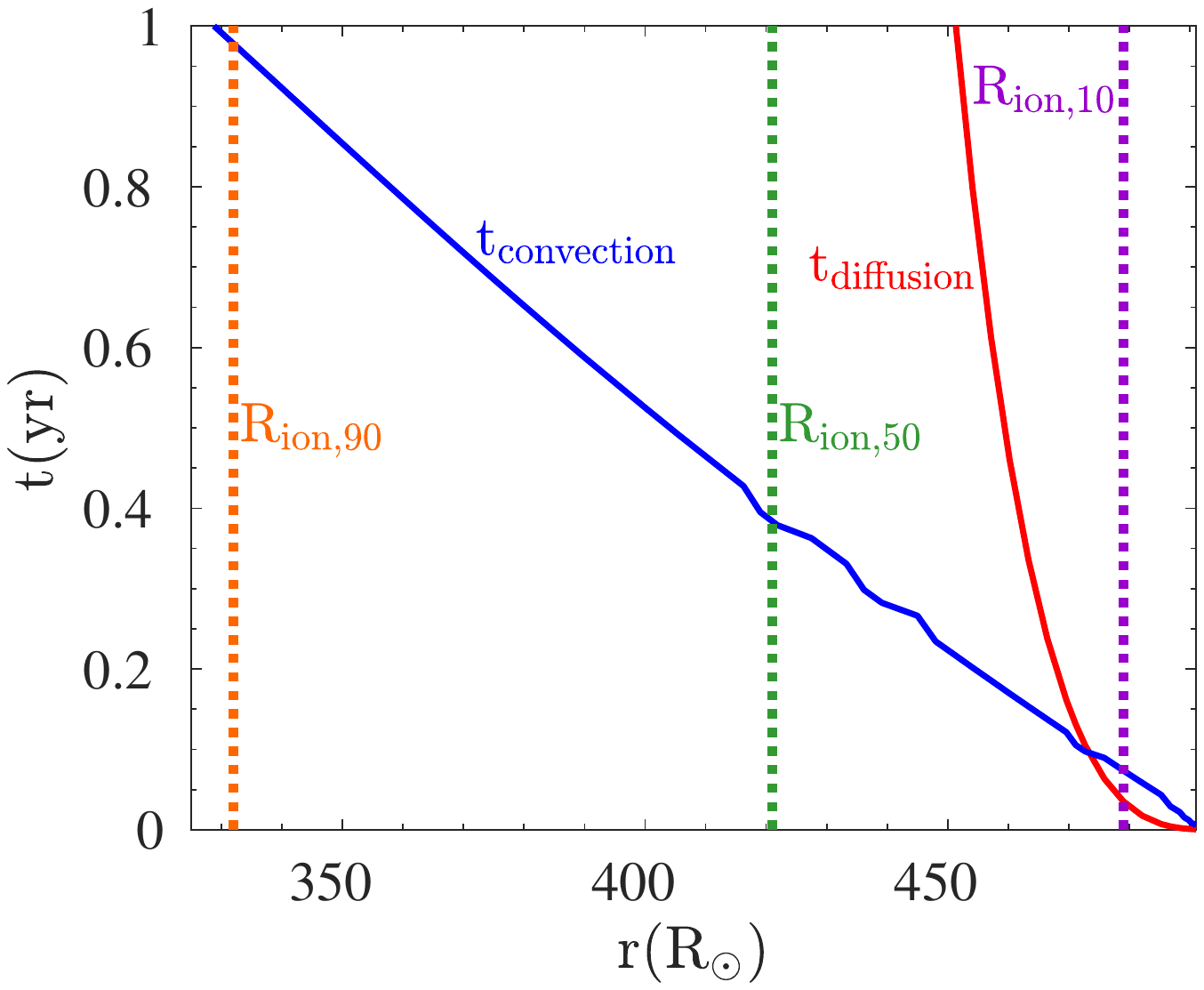}
\vspace*{-4.5cm}
\caption{
Like Fig. \ref{fig:DiffConv0}  but at the end of the plunge-in phase at  $t=1.7 \yr$. At this time $t_{\rm diffusion}({\rm H}^+) = 5.9 \yr$.
}
\label{fig:DiffConv1.7}
\end{center}
\end{figure}

From the figures we learn the following about the energy transport times in the inflated envelope. 
\begin{enumerate}
  \item Convection can carry energy out from the entire hydrogen recombination zone very efficiently, as the convection transport time from $90\%$ hydrogen ionization fraction to the surface is shorter than the plunge-in time, $1.7 \yr$ in our case.   
 \item The convection cells can move even faster, up to the sound speed that is larger than the convective speed of the models at the different times (Figs. \ref{fig:rhoTV0}-\ref{fig:rhoTV17})
 \item In the outer region photon diffusion time is shorter than the convection time. For example, at $t=1.7 \yr$ the energy can be carried out from $R_{\rm ion,50}$, where hydrogen is $50 \%$ ionized, to $r=480 R_\odot$ in about $0.3 \yr$, much shorter than the plunge-in time. From there the photon diffusion time is very short and photons carry the energy very rapidly out.  
 \item Even the photon diffusion alone cannot be neglected. The photon diffusion  times from $R_{\rm ion,50}$ at the different times are 
$t_{\rm diffusion}({\rm H}^+, 0) = 2.2 \yr$, 
$t_{\rm diffusion}({\rm H}^+,0.8) = 3.5$,  and
$t_{\rm diffusion}({\rm H}^+, 1.7) = 5.9 \yr$, respectively. These times are not much longer than the plunge in time. As we found, convection is more important than radiation. 
 \item The transport time from the respective ionization fraction of hydrogen zones increases as the envelope is inflated. However, our calculation does not include mass removal. We expect that mass removal will reduce the optical depth and will reduce the photon diffusion time.   
\end{enumerate}

The above findings hint on the outcome of a larger energy deposition. Consider a case where the companion continues the plunge-in phase to inner envelope layers, say as deep as $15 R_\odot$. It would release then about 3 times more energy. As convection velocity can reach the sound speed, the convection is capable of adjusting itself to the energy increase. From Fig. \ref{fig:rhoTV17} we see that the convection speed can increase up to the sound speed by about a factor of $f_{\rm conv,10}\equiv c_s/v_{\rm conv} =1.7$, $f_{\rm conv,50}=2.6$, and $f_{\rm conv,90}=4.1$ at the end of the calculated plunge-in phase. The number in the subscript indicates that the factor is taken at the radius where the percentage of ionized hydrogen has this value. We performed the integration according to equation \ref{eq:Tout} and found at the end of the plunge-in phase ($t=1.7 \yr$) that where hydrogen is 90 per cent ionized the minimum convection energy transport time out can be as short as $t_{\rm conv,min,90}=0.36 \yr$. This is about 3 times shorter than $t_{\rm convection} =1 \yr$ at the same radius in the model presented in Fig. \ref {fig:DiffConv1.7}. The same arguments bring us to conclude that convection could carry the recombination energy out even if the plunge-in time to a radius of $50 R_\odot$ would be shorter by a factor of 3 from what we take in our simulations, e.g., about 7 months instead of 20 months ($1.7 \yr$). It cannot be shorter due to the limitation of dynamical in-spiral-time, and hence we consider it unlikely for the plunge-in time be shorter than $\approx10$ months.  
  
We conclude that during the plunge-in phase convection can carry as much as $\approx 3$ times more energy than in our simulations in the present study. This would be the case, as we indicated above, if the same companion spirals-in much deeper in the plunge-in phase, down to $\simeq 15 R_\odot$. However, a more accurate result than our simple estimate requires more accurate calculations than what we can perform with MESA. Such calculations must take into account the response of deeper envelope layers as well as some mass ejection from the outer envelope, and hence dynamical effects become important for such a deep plunge-in, all in a flatten envelope.  

We can phrase our main conclusion regarding recombination energy transport during envelope inflation as follows. 
Although the amount of mass outward to the ionization zone of hydrogen increases and so does the energy transport time, convection and photon diffusion are efficient enough to carry most of the hydrogen recombination energy out.  

\section{SUMMARY}
\label{sec:summary}

There is an ongoing two decade old dispute on the importance of the recombination energy of hydrogen and helium in assisting the removal of the common envelope, with claims for its high importance (e.g., \citealt{Hanetal1994, Hanetal2002}) and claims for its limited role (e.g., \citealt{Harpaz1998, SokerHarpaz2003}), to list old papers on the subject (see section \ref{sec:intro} for recent papers).  
 
 In a recent paper \cite{Sabachetal2017} examined the transport of energy in the envelope of an AGB star both by convection and by photon diffusion. They concluded that the energy transport in the envelope substantially reduces the fraction of the recombination energy of hydrogen and helium that is available for envelope removal. They, however, did not follow the inflation of the envelope in the early phase of the CEE when the secondary star rapidly spirals-in, i.e., the plunge in phase. We set here the goal of examining the variation of the energy transport time as the envelope expands.
 
We presented our main results in Figs. \ref{fig:DiffConv0}-\ref{fig:DiffConv1.7}. We found that as the envelope expands both the convection time and photon diffusion time increase, but both stay below the evolution time (in the relevant parts of the envelope).
The convection is capable of carrying most of the recombination energy to the very outer part of the envelope, where radiation becomes important and photons carry the energy out of the star.
 
We emphasize that the convective times we calculated hold only for the convective regions of the envelope. In envelope regions that are already flowing out at a relatively high speed, which we did not study here, there is no hydrostatic equilibrium anymore and there is no convection. However, in an envelope that expands at a relatively high speed, about and above the escape speed, the optical depth will be low and photon diffusion time is expected to be short.
 
We limited ourselves to the recombination of hydrogen and to the plunge-in phase. To study the recombination energy of helium and the later evolution of the envelope there is a need to use more sophisticated numerical codes that include the non-spherical structure of the envelope. Such a structure results from the rotation and concentration of mass ejection toward the equatorial plane. 

Our findings are good news for observations as we conclude that the recombination energy will be radiated away, hence making it possible to observe the early phases of the CEE. Another source of radiated energy can be the accretion of mass onto the secondary star. Such an event of high luminosity over weeks (when the primary  star is small) to years might be classified as an intermediate luminosity optical transient (ILOT) event, that is expected at the beginning of the CEE (e.g. \citealt{RetterMarom2003, Retteretal2006, Tylendaetal2011, Tylendaetal2013, Nandezetal2014, Zhuetal2016, Soker2016GEE, Galavizetal2017, MacLeodetal2017M31, MacLeodetal2018}). 

 We can summarize our main result by stating that one cannot include the recombination energy in the study of the common envelope evolution without considering convection and radiative transfer. Adding the recombination energy to the total energy budget, e.g., in the equation of state, in numerical simulations that do not include energy transport might lead to inaccurate results. 

\section*{Acknowledgments}
We thank an anonymous referee for comments that improved the presentation of our results. We acknowledge support from the Israel Science Foundation and a grant from the Asher Space Research Institute at the Technion. A.G. was supported by The Rothschild Scholars Program- Technion Program for Excellence. 





\label{lastpage}
\end{document}